\documentclass[3p,11pt,twocolumn,sort&compress]{elsarticle}
\usepackage{amsfonts}
\usepackage{amsmath}

\usepackage{lineno}
\usepackage{xcolor}
\usepackage{todonotes}
\usepackage{xspace}

\newcommand{\diff}{\ensuremath{\mathrm{d}}}
\newcommand{\plab}{\ensuremath{p_\text{lab}}\xspace}
\newcommand{\elab}{\ensuremath{E_\text{lab}}\xspace}

\journal{Physics Letters  B}
\usepackage{hyperref}
\begin{document}
%\linenumbers
\begin{frontmatter}



\title{On the $\eta$ and $\eta'$ Photoproduction Beam Asymmetry at High Energies}

\author[CEEM,IU]{V.~Mathieu \corref{cor1}}
\ead{mathieuv@indiana.edu}
\author[ghent]{J.~Nys}
\author[UNAM]{C.~Fern\'andez-Ram\'irez}
\author[CEEM,IU]{A.~Jackura}
\author[bonn]{M.~Mikhasenko}
\author[jlab]{A.~Pilloni}
\author[CEEM,IU,jlab]{A.~P.~Szczepaniak}
\author[IUinfo]{G. Fox}
\address[CEEM]{Center for Exploration of Energy and Matter, Indiana University, Bloomington, IN 47403, USA}
\address[IU]{Physics Department, Indiana University, Bloomington, IN 47405, USA}
\address[ghent]{Department of Physics and Astronomy, Ghent University, B-9000 Ghent, Belgium}
\address[UNAM]{Instituto de Ciencias Nucleares, Universidad Nacional Aut\'onoma de M\'exico, Ciudad de M\'exico 04510, Mexico}
\address[bonn]{Universit\"at Bonn, Helmholtz-Institut f\"ur Strahlen- und Kernphysik, 53115 Bonn, Germany}
\address[jlab]{Theory Center, Thomas Jefferson National Accelerator Facility,12000 Jefferson Avenue, Newport News, VA 23606, USA}
\address[IUinfo]{School of Informatics and Computing, Indiana University, Bloomington, IN 47405, USA}

\author{\\[.4cm] (Joint Physics Analysis Center)}

\cortext[cor1]{Corresponding author}

\begin{abstract}
We show that, in the Regge limit, beam asymmetries in $\eta$ and $\eta'$ photoproduction are sensitive to hidden strangeness components. Under reasonable assumptions about the couplings we estimate the contribution of the $\phi$ Regge pole, which is expected to be the dominant hidden strangeness contribution. The ratio of the asymmetries in $\eta'$ and $\eta$ production is estimated to be close to unity in the forward region $0 < -t/\text{GeV}^2 \leq 1$ at the photon energy  $\elab = 9$~GeV, relevant for the upcoming measurements at Jefferson Lab. 
\end{abstract}

\begin{keyword}
\PACS 11.55.Jy \sep 12.39.Jh\\ 
JLAB-THY-17-2450
\end{keyword}

\end{frontmatter}

Meson photoproduction plays an important role in studies of the hadron spectrum and searches for exotic mesons~\cite{Close:1994pr,Afanasev:1999rb,Szczepaniak:2001qz,Meyer:2015eta,physrept}, in particular for hybrids. The latter contain a large gluon component and are predicted in phenomenological models of QCD and lattice simulations~\cite{Isgur:1984bm,Swanson:1997wy,Buisseret:2006sz,McNeile:2006bz,Dudek:2011bn,Dudek:2013yja}. Identifying the nature of new resonances requires to establish their quantum numbers first, which constrain both decays and production mechanisms. At Jefferson Lab, with photon energies $\elab \sim 5-11$ GeV, meson resonances are produced via beam fragmentation, which is expected to be dominated by exchanges of leading Regge poles~\cite{Battaglieri:2014gca}. 

Production of the lightest multiplet of exotic mesons with $J^{PC} = 1^{-+}$ involves the same Regge exchanges that appear in production of ordinary pseudoscalar mesons, like $\pi^0$, $\eta$ and $\eta'$, and both natural ($P(-1)^J =1$) and unnatural ($P(-1)^J=-1$) parity exchanges contribute. 
One of the key observables which is sensitive to the exchange process is the beam asymmetry. It is related to the ratio of cross sections for natural and unnatural Regge exchanges and yields precise information about the resonance production mechanism. The GlueX experiment recently measured $\pi^0$ and $\eta$ beam asymmetries~\cite{AlGhoul:2017nbp} and the measurement of $\eta'$ is expected soon. Similar measurements will also be performed by the CLAS12 experiments~\cite{battaglieri2005meson} in the near future.

In this letter we give an estimate for the $\eta'$ photoproduction beam asymmetry at high energies. The beam asymmetry is defined as 
\begin{align}
\Sigma^{(\prime)} & = \frac{\diff \sigma^{(\prime)}_\perp -  \diff \sigma^{(\prime)}_\parallel}{\diff \sigma^{(\prime)}_\perp +  \diff \sigma^{(\prime)}_\parallel},
\end{align}
with $\diff \sigma_{\perp,\parallel} \equiv  \frac{\diff\sigma_{\perp,\parallel}}{\diff t} (s,t)$ denoting the differential cross section for photons polarized perpendicular or parallel to the reaction plane, $s$ and $t$ are the usual Mandelstam variables.
Unprimed and primed quantities refer to $\eta$ and $\eta'$, respectively. We wish to estimate the quantity 
\begin{align} \nonumber
\frac{\Sigma'}{\Sigma}  & = 1+
\frac{ 2( \diff\sigma'_\perp \diff\sigma_\parallel   -   \diff\sigma_\perp \diff\sigma'_\parallel)}
{ (\diff\sigma'_\perp+\diff\sigma'_\parallel ) (\diff\sigma_\perp - \diff\sigma_\parallel)} \\ \label{eq:epsilon} 
& \equiv 1 + \epsilon.
\end{align}
Using the recent measurements of the $\eta$ beam asymmetry \cite{AlGhoul:2017nbp,Nys:2016vjz}, one can extract the quantities
\begin{align}
\frac{2\diff \sigma_\perp}{\diff \sigma_\perp+\diff \sigma_\parallel} & =1+\Sigma,&
\frac{2\diff \sigma_\parallel}{\diff \sigma_\perp+\diff \sigma_\parallel} & = 1-\Sigma.
\end{align}
It is convenient to rewrite the ratio under interest as 
\begin{align} \label{eq:ratio}
\frac{\Sigma'}{\Sigma}  & = 1+ \frac{1-\Sigma^2}{\Sigma} \cdot 
\frac{k_V - k_A}{(1+\Sigma) k_V + (1-\Sigma) k_A}\,.
\end{align}
In order to evaluate this ratio, one must determine the quantities
\begin{align}\label{eq:kNkU}
k_V &= \frac{\diff\sigma'_\perp}{\diff\sigma_\perp}, &
k_A &= \frac{\diff\sigma'_\parallel}{\diff\sigma_\parallel}.
\end{align}
In our evaluation of $k_{V,A}$, we proceed as follows. We first identify the Regge 
poles. For the natural exchanges, we extract their residues at the photon vertex by considering the radiative decays and the residues at the nucleon vertex from nucleon-nucleon total cross section. We then estimate the residues of the unnatural exchanges. Finally we give our prediction for $\epsilon$ in Eq.~\eqref{eq:epsilon} and list the possible deviations from our assumptions.

Regge poles are classified according to the internal quantum numbers of the particles with the lowest spin located on the corresponding trajectory. The natural exchanges dominating the $\eta^{(\prime)}$ photoproduction are $\rho$, $\omega$ and $\phi$, and the unnatural ones are $b$, $h$ and $h'$.\footnote{The lowest spin exchange in the $b$ trajectory is  the isovector $1^{+-}$ state in the PDG~\cite{pdg}, the $b_1(1235)$. Similarly, the lowest spin exchange on the $h$ and $h'$ trajectory are the isoscalar $1^{+-}$ states, the $h_1(1170)$ and $h_1(1380)$. The former is observed decaying into $\rho\pi$, the latter into $\bar K K^*(892)$, which suggests ideal mixing.}
Asymptotically 
$\diff\sigma_\perp$ and $\diff\sigma_\parallel$ are dominated by natural and unnatural exchanges, respectively, and so are  the corresponding cross section ratios $k_V$ and $k_A$. 
In absence of hidden strangeness ($\bar s s$) in the proton and for a vanishing contribution from the associated exchange of $\phi$ and $h'$ mesons,\footnote{We assume ideal $\omega/\phi$ and $h/h'$ mixing.} one finds 
\begin{equation} 
k_V = k_A = \tan^2\varphi, \label{kk} 
\end{equation}
where $\varphi$ is the $\eta-\eta'$ mixing angle in the flavor basis. %Phenomenological analyses of radiative decays give $ \tan \varphi\sim 0.7-0.9$~\cite{Bramon:2000fr}, consistent with predictions of chiral Lagrangians~\cite{Degrande:2009ps,Mathieu:2010ss}.
Analysis of $\eta/\eta'$ transition form factors 
 gives $\phi = 39.3^o(1.2)$~\cite{Escribano:2015nra}, which is somewhat different from the prediction $\phi = 41.4^o$ based on chiral Lagrangians~\cite{Mathieu:2010ss}.
Eq.~\eqref{kk} implies $\Sigma' = \Sigma$, so one concludes that a sizable deviation of the ratio of beam asymmetries from unity indicates either non-negligible contributions from hidden strangeness $\phi$ and $h'$ exchanges, and/or significant deviation from the quark model description of the $\eta$ mesons or from the Regge pole dominance. 

The leading contributions to the differential cross section are\footnote{In~\cite{Mathieu:2015eia} we numerically showed that this approximation is valid for $\elab >5$ GeV in the forward direction.} 
\begin{subequations}\label{eq:cgln}
\begin{align} 
\frac{\diff\sigma^{(\prime)}_\perp}{\diff t}(s,t) & =K\left( |A^{(\prime)}_1|^2 - t|A^{(\prime)}_4|^2 \right), \\ 
\frac{\diff\sigma^{(\prime)}_\parallel}{\diff t}(s,t) & = K \left(|A^{(\prime)}_1 +t A^{(\prime)}_2|^2 - t|A^{(\prime)}_3|^2 \right),
\end{align}
\end{subequations}
where $K$ is a kinematical factor that cancels out in the polarization observables. The $A_i$, $i=1,\ldots, 4$ are the conventional CGLN invariant amplitudes \cite{Chew:1957tf}. At leading order in the energy, the scalar amplitudes $A_i$ are related to the $s-$channel helicity amplitudes $A_{\lambda', \lambda \lambda_\gamma}$\footnote{$\lambda,\lambda'$ and $\lambda_\gamma$ are the $s-$channel helicities of the target, the recoil and the beam respectively. We denote by $\pm$ the nucleon helicities $\pm \frac{1}{2}$ for brevity.} by
\begin{subequations}\label{eq:PCH}
\begin{align}
A_{+,+1} + A_{-,-1} & = \sqrt{2} s \sqrt{-t} A_3 \,,\\
A_{+,+1} - A_{-,-1} & = \sqrt{2} s \sqrt{-t} A_4 \,,\\
A_{-,+1} + A_{+,-1} & = -\sqrt{2} s (A_1 + t A_2) \,,\\
A_{-,+1} - A_{+,-1} & = -\sqrt{2} s A_1\,.
\end{align}
\end{subequations}
These combinations of ($s-$channel) helicity amplitudes are parity conserving in the $t-$channel. $A_1$ and $A_4$ involve natural exchanges and $A_1+tA_2$ and $A_3$ involve unnatural exchanges.\footnote{See Appendix A of~\cite{Mathieu:2015eia} for a detailed discussion on the quantum numbers corresponding to the invariant amplitudes $A_i$.} 
Specifically, we write the scalar amplitudes $A_i = \sum_V A_i^V + \sum_A A_i^A$ with the natural Regge poles $V = \rho,\omega,\phi$ and the unnatural Regge poles $A = b,h,h'$. The natural exchanges are (with $s$ in units of GeV$^2$)
\begin{subequations}\label{eq:A1A4}
\begin{align} 
A_{1,4}^{(\prime)V}(s,t) & =  \beta^{(\prime)V}_{1,4}(t) \frac{1- e^{-i\pi \alpha_V(t)}}{ \sin\pi \alpha_V(t)} s^{\alpha_V(t)-1}, \\ 
A^{(\prime)V}_2(s,t) & = (-1/t) A^{(\prime)V}_1, \\ 
A^{(\prime)V}_3(s,t) & = 0.
\end{align}
\end{subequations}
The factorization of Regge residues yields a simple form for the kinematical singularities in $t$~\cite{Cohen-Tannoudji:1968eoa}
\begin{align} \label{eq:kinT}
A_{\lambda', \lambda \lambda_\gamma} & \propto
\left(\sqrt{-t}\right)^{|\lambda_\gamma| + |\lambda-\lambda'|}.
\end{align}
Comparing Eq.~\eqref{eq:PCH} and Eq.~\eqref{eq:kinT}, we see that $A_1$, hence the residue $\beta_1(t)$, needs to be proportional to $t$.
In the (physical) region under consideration, {\it i.e.} the forward direction where $t$ is small and negative, the residues $\beta^{(\prime)V}_i(t)$, with kinematical singularities removed, are regular real functions of the momentum transfered $t$. A standard parametrization of the residues~\cite{Irving:1977ea,Worden:1972dc,Nys:2016vjz}, $\beta \propto 1/\Gamma(\alpha)$ describes both the exponential suppression seen in data and zeros at ghost poles.\footnote{The ghost poles are the poles when $\alpha$ is a negative integer.} In general, however, since all natural (unnatural) poles have approximatively the same trajectory $\alpha_V(t)$ ($\alpha_A(t)$), the beam asymmetry depends weakly on the details of the $t$-dependence ({\it c.f.} Eq.~\eqref{eq:epsilon})  and $\epsilon$ is primarily determined by the relative strengths of the various exchanges. Accordingly, for the evaluation of the ratio in Eq.~\eqref{eq:epsilon}, we use
\begin{equation}
\begin{split}
\beta^{(\prime) V}_1(t) & = g^{(\prime)}_{VP\gamma} g_{1V} t  e^{b t}, \\
\beta^{(\prime) V}_4(t) & = g^{(\prime)}_{VP\gamma} g_{4V} e^{b t}.  
\end{split}\label{eq:betas}
\end{equation}
The additional factor of $t$ in $\beta_1^V(t)$ is required by factorization of Regge residues, as we noticed above, and can be understood using an effective Lagrangian to desribe the exchange of a $\rho$ meson~\cite{Mathieu:2015eia}. The  $g_{1 V}$ and $g_{4 V}$ denote the nucleon couplings and the $g_{VP\gamma}^{(\prime)}$ denotes the coupling at the $\gamma \eta^{(\prime)}$ vertex. In Eq.~\eqref{eq:betas}, we kept a universal exponential slope $b$. In the ratio of beam asymmetries in Eq.~\eqref{eq:epsilon} the exponential factor cancels out. It is needed for the determination of the $\rho$ nucleon helicity flip coupling $g_{1 \rho}$ when fitting $\pi^- p \to \pi^0 n$ differential cross section, discussed below. 

It is well known that the $\rho$ and $\omega$ trajectories are almost degenerate $\alpha_\omega(t)=\alpha_{\rho}(t) = 0.9t+0.5$ (with $t$ expressed in GeV$^2$). For the $\phi$ Reggeon we assume the same slope ($\alpha'$), but take into account the difference between the masses that determine the intercepts $\alpha_\omega(0)-\alpha_\phi(0) = \alpha'(m_\phi^2-m_\omega^2) \sim 0.5$, so that $\alpha_{\phi}(t) = 0.9t$. We define  
\begin{align}
r(t) = \frac{1-e^{-i \pi \alpha_\phi(t)}}{1-e^{-i \pi \alpha_\omega(t)}} \frac{\sin \pi \alpha_\omega(t)}{\sin \pi \alpha_\phi(t)} s^{\alpha_\phi(t)-\alpha_\omega(t)}
\end{align}
and, with the amplitudes described above, one obtains, 
\begin{align}\label{eq:kNdef}
k_V  &= 
\Big(\left| g'_{\rho \gamma} g_{4 \rho} + g'_{\omega \gamma} g_{4 \omega} + r(t) g'_{\phi \gamma} g_{4 \phi} \right|^2 \nonumber\\
&\quad- t \left| g'_{\rho \gamma} g_{1 \rho} + g'_{\omega \gamma} g_{1 \omega} + r(t) g'_{\phi \gamma} g_{1 \phi} \right|^2 \Big)\Big/ \nonumber\\
&\quad\Big(\left| g_{\rho \gamma} g_{4 \rho} + g_{\omega \gamma} g_{4 \omega} + r(t) g_{\phi \gamma} g_{4 \phi} \right|^2 \nonumber\\
&\quad- t \left| g_{\rho \gamma} g_{1 \rho} + g_{\omega \gamma} g_{1 \omega} + r(t) g_{\phi \gamma} g_{1 \phi} \right|^2 \Big).
\end{align}

Factorization of Regge residues allowed us to write the residues as a product 
of $\gamma\eta^{(\prime)}$ coupling $g^{(\prime)}_{VP\gamma}$ and two nucleon couplings ($g_{1V}$ and $g_{4V}$).  The photon couplings in $\eta$ and $\eta'$ photoproduction can be estimated from radiative decays  using
\begin{subequations}
\begin{align}
\Gamma (V \to \gamma P) & = \frac{g_{VP\gamma}^2}{12 \pi} \left( \frac{m_V^2-m_P^2}{2 m_V} \right)^3, \\
\Gamma (P \to \gamma V) & = \frac{g_{VP\gamma}^2}{4\pi} \left( \frac{m_P^2-m_V^2}{2 m_P} \right)^3.
\end{align}
\end{subequations}
The extracted couplings from \cite{pdg} are summarized in Table~\ref{tab:1}.
\begin{table}[htb]\caption{Radiative decays and extracted couplings.\label{tab:1}}\begin{center}
    \begin{tabular}{|c|c|c|}
    \hline
    ~ & $\Gamma$ (keV) & $g_{VP\gamma}$ (GeV$^{-1}$) \\ \hline
    $\rho \to \eta \gamma$ & 44.7(3.1) & 0.480(17) \\
    $\eta' \to \rho \gamma$ & 56.6(2.8) & 0.398(10) \\
    $\omega \to \eta \gamma$ & 3.82(34) & 0.135(6) \\
    $\eta' \to \omega \gamma$ & 5.14(35) & 0.127(4) \\
    $\phi \to \eta \gamma$ & 55.6(1.0) & 0.210(2) \\
    $\phi \to \eta' \gamma$ & 0.265(9) & 0.216(4) \\
    \hline
    \end{tabular}
\end{center}
\end{table}

\begin{figure}[htb]
\centerline{
\includegraphics[width=\linewidth]{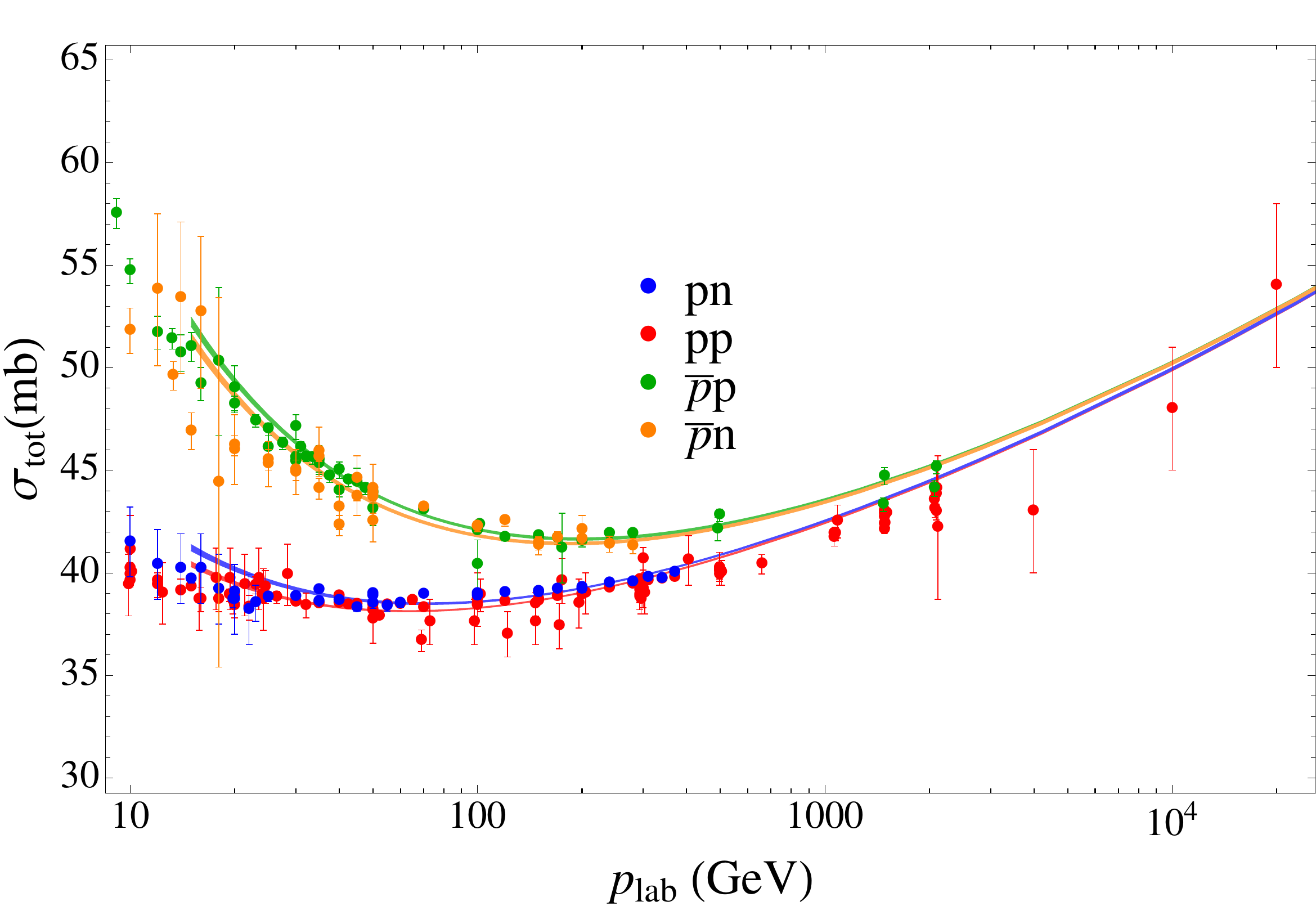}}
\caption{\label{fig:sigtot}Total cross section. Fit compared to data from PDG~\cite{pdg}.}
\end{figure}

\begin{figure}[htb]
\centerline{
\includegraphics[width=0.95\linewidth]{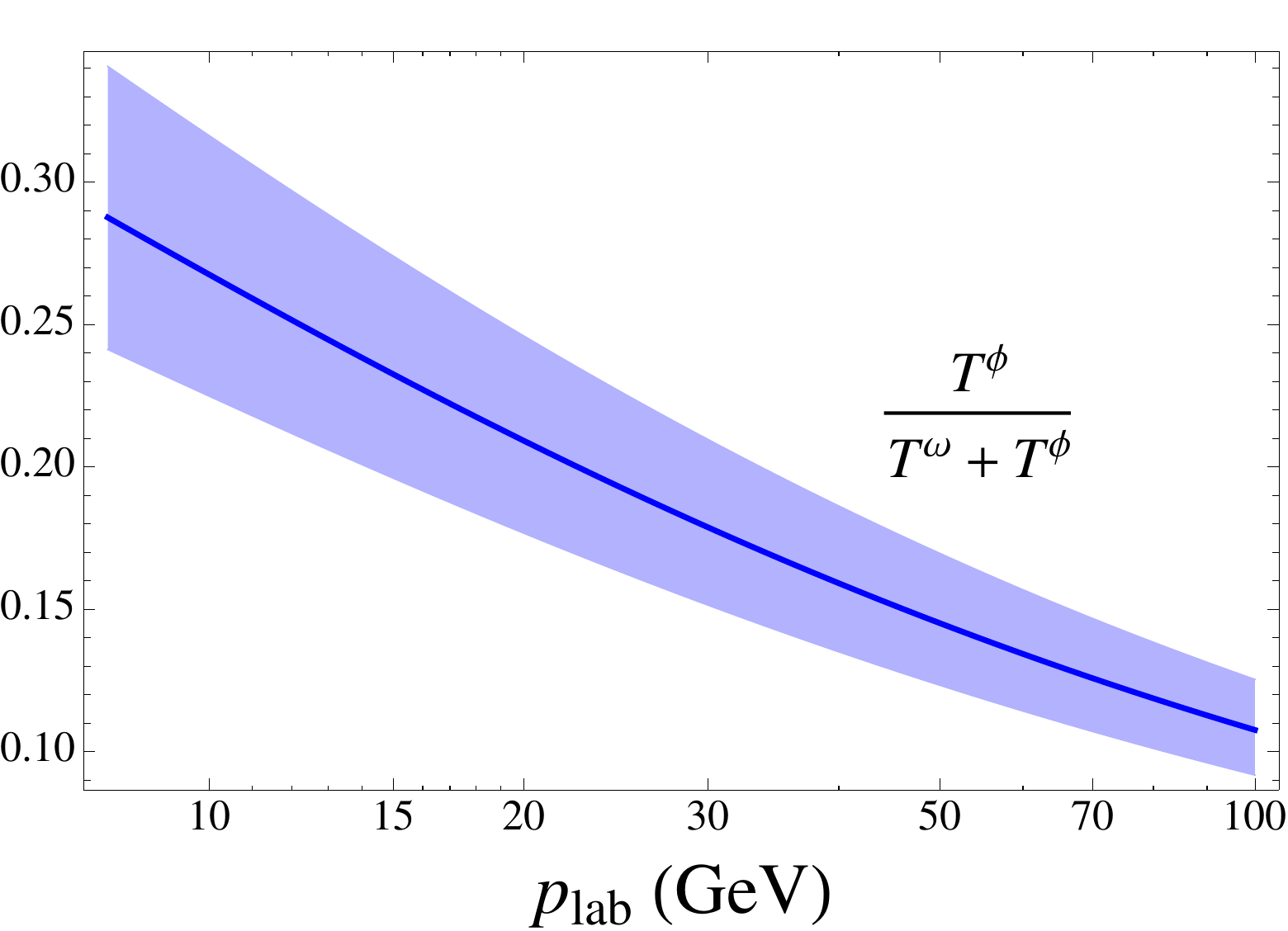}}
\caption{\label{fig:ratio}Relative strength of the hidden strangeness exchange to the total cross section.}
\end{figure}

At leading order in the energy,  $g_{1V}$ and $g_{4V}$ correspond to the $s$-channel helicity flip and non-flip couplings at the nucleon vertex respectively~\cite{Mathieu:2015eia}. By denoting the flip and non-flip amplitudes 
%via the natural Regge pole $V$ 
as $A^V_{+, \pm 1}$, in the high-energy limit one obtains
\begin{align} \label{eq:g14}
\frac{g_{1V}}{g_{4V}} & = \frac{ A^V_{+,-1} }{A^V_{+,+1}} \frac{1}{\sqrt{-t}}.
\end{align}
The nucleon non-flip couplings $g_{4V}$ are determined by fitting the $pp$, $\bar p p$, $pn$ and $\bar p n$ total cross sections. At high energies the relevant exchanges contributing to these are the Pomeron $\mathbb P$ and the Regge poles $f_2$, $a_2$, $\rho$, $\omega$ and $\phi$. From the optical theorem it follows that the total cross section is related to imaginary part of the forward scattering amplitude.
We denote $T^V(s) \equiv \text{Im }A^V_{NN\to NN}(s,t=0) = g_{4 V}^2 s^{\alpha_0^V}$. The axial exchanges vanish in the forward direction because of charge conjugation invariance, and do not contribute to the total cross section.  
The relative contribution of individual poles to the total cross section is given by
\begin{subequations}
\begin{align} \nonumber
\sigma_{\text{tot}}^{\overset{(-)}{p} p}(s)  &=  \frac{1}{4 \plab m_N} \Big[ T^{\mathbb P}(s)  + T^{f_2}(s) - T^{a_2}(s) \\
& \quad\mp  T^{\omega}(s)  \mp T^{\phi}(s) \mp T^{\rho}(s)  \Big],
\\ \nonumber
\sigma_{\text{tot}}^{\overset{(-)}{p} n}(s) &= \frac{1}{4 \plab m_N} \Big[ T^{\mathbb P}(s)  + T^{f_2}(s) + T^{a_2}(s) \\ 
&\quad \mp T^{\omega}(s)  \mp T^{\phi}(s) \pm T^{\rho}(s)   \Big],
\end{align}
\end{subequations}
where the top (bottom) sign corresponds to a (anti-)proton beam. 
We use the intercept values $\alpha_0^{\rho} = \alpha_0^{\omega} = \alpha_0^{a_2}  = \alpha_0^{f_2}  =0.5$, $\alpha_0^\phi = 0.0$ and $\alpha_0^{\mathbb P} = 1.08$. Only the data with $p_\text{lab}\geq 15$ GeV are included in the fit. The relevant couplings resulting from the fit are summarized in Table~\ref{tab:2}. The comparison to the data is presented on Fig.~\ref{fig:sigtot}. We note that our value for the ratio of the nucleon couplings $g_{4\phi}/g_{4\omega} = 1.29(9)$ is significantly bigger than the same ratio extracted in other analyses. For instance in~\cite{Williams:1996bd} it was found $g_{4\phi}/g_{4\omega} = 0.34$. This is because we neglected a large systematic uncertainty on $g_{4\phi}$. We indeed note that $g_{4\phi}$ depends on the data selected for the fit. For instance if instead of selecting data with $p_\text{lab}> 15$ GeV, we choose $p_\text{lab}> 30$ GeV we obtain $g_{4\phi}=  4.20(1.72)$ where the other nucleon couplings $g_{4\rho}$ and $g_{4\omega}$ remain unchanged. In our approach the influence of the $\phi$ is nevertheless suppressed compared to the $\omega$ exchange by the difference in their intercept. This is evident on Fig.~\ref{fig:ratio} where we illustrate the relative strength of the hidden strangeness exchange by the ratio $T^\phi(s)/(T^\omega(s) + T^{\phi}(s))$. Our value for this ratio lies in the range $0.1 - 0.3$ in the region $\plab =  10 - 100$ GeV (see Fig~\ref{fig:sigtot}).  
It is worth noticing that the coupling of the nucleon to the $\phi$ meson can be related to the strange electromagnetic form factors $G_E^s$ and $G_M^s$~\cite{Mergell:1995bf,Belushkin:2006qa,Pacetti:2015iqa}. These can either extracted from lattice simulations~\cite{Green:2015wqa,Shanahan:2014tja}, or inferred by measurement of low-energy parity violation in $\vec{e}p$ scattering~\cite{Kaplan:1988ku,Alberico:2001sd,Beise:2004py,Acha:2006my,Androic:2009aa,Baunack:2009gy,Ahmed:2011vp,GonzalezJimenez:2011fq,vanOers:2007if,JLABexpPV}. This might be an example of how high energy measurements can help in constraining the low energy information.

The isoscalar exchanges are empirically found to be dominated by non-flip at the nucleon vertex~\cite{Irving:1977ea}. Accordingly, we set $g_{1 \phi} = g_{1 \omega} = 0$. We therefore only need to determine the nucleon helicity-flip coupling of the $\rho$ exchange. To this end, we analyze high-energy $\pi^- p \to \pi^0 n$ data (which contains contribution from the $\rho$ pole only) using
\begin{subequations} \label{eq:CEX}
\begin{align}
A_{++}  &= g_{\rho\pi}g_{4 \rho} e^{bt} \frac{1-e^{-i\pi \alpha_\rho(t)}}{\sin \pi \alpha_\rho(t)} s^{\alpha_\rho(t)}, \\
A_{+-}  & = g_{\rho\pi} g_{1 \rho } e^{bt}  \frac{1-e^{-i\pi \alpha_\rho(t)}}{\sin \pi \alpha_\rho(t)} s^{\alpha_\rho(t)} \sqrt{-t} .
\end{align}	
\end{subequations}
We fit the high-energy data of ~\cite{Barnes:1976ek} in the forward direction $|t|\leq 0.2$ GeV$^2$. The differential cross section is given by
\begin{align}
\frac{\diff \sigma}{\diff t}(s,t) & = \frac{1}{64\pi m_N^2 \plab^2} \left ( \left| A_{++} \right|^2 + \left| A_{+-} \right|^2 \right).
\end{align}
We do not attempt to fit larger values of $|t|$ since our model has a very simple $t$ dependence. 

\begin{figure}[htb]
\centerline{\includegraphics[width=\linewidth]{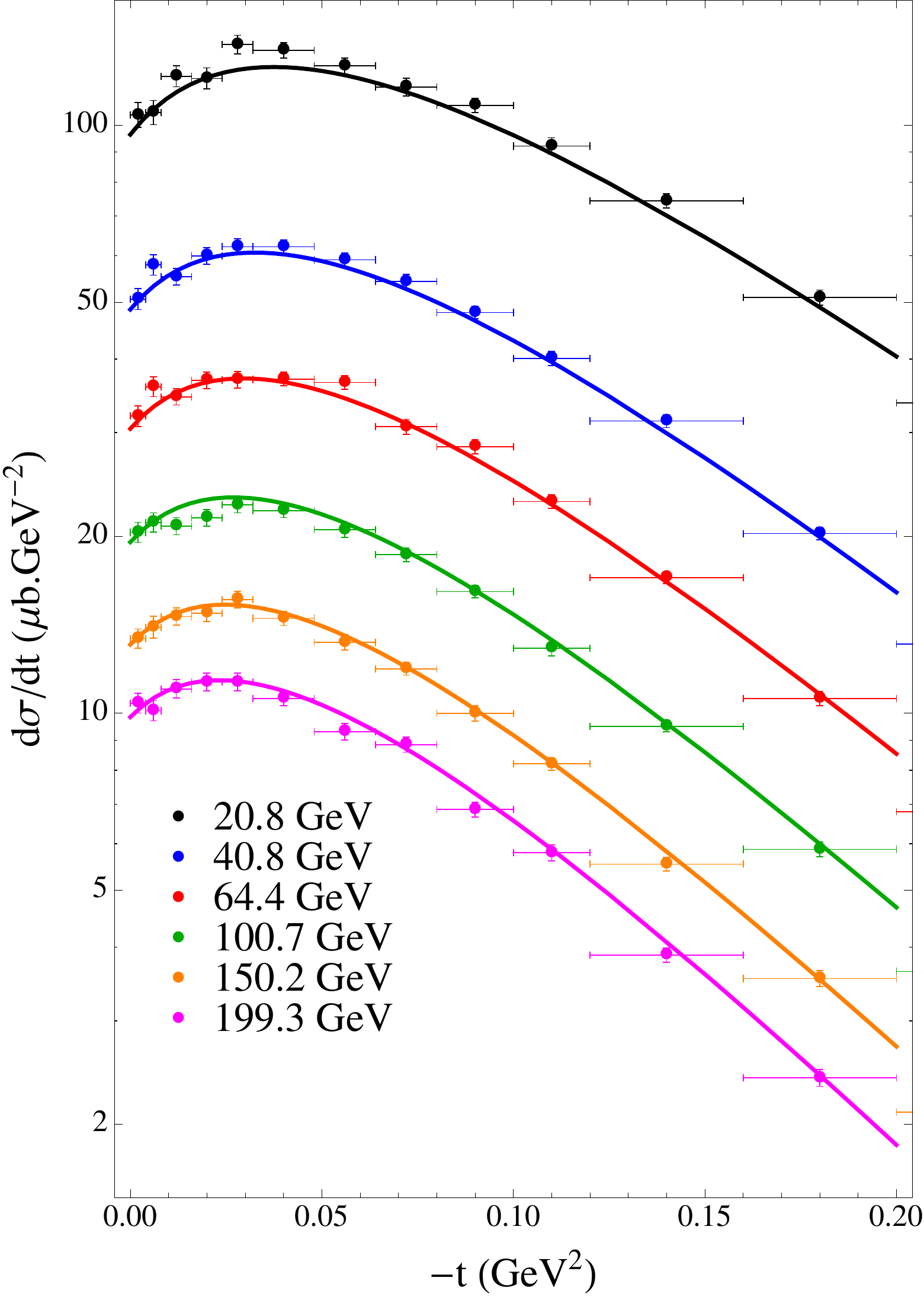}}
\caption{\label{fig:CEX}$\pi^- p \to \pi^0 n$ differential cross section at high energies in the forward direction. Comparison of the model with data from~\cite{Barnes:1976ek}.}
\end{figure}

The main purpose of this fit is to estimate the nucleon helicity-flip coupling $g_{1\rho}$. The normalization of the ratio of the amplitudes in Eqs.~\eqref{eq:CEX} has been chosen to be in agreement with Eq.~\eqref{eq:g14}. The fit yields $b = 2.97(7) \text{ GeV}^{-2}$ and $g_{1\rho}/g_{4\rho} = 5.91(7) \text{ GeV}^{-1}$ (in agreement with the standard value $g_{1\rho}/g_{4\rho} \simeq 6\text{ GeV}^{-1}$~\cite{Irving:1977ea}) from which one obtains $g_{1\rho} = 13.59(45) \text{ GeV}^{-3}$. The comparison of the model with data is shown in Fig.~\ref{fig:CEX}.
\begin{table}[htb]\caption{Nucleon couplings.\label{tab:2}}\begin{center}
    \begin{tabular}{|c|c|c|}
    \hline
    ~ & $g_{1V}$ ($\text{GeV}^{-3}$) & $g_{4V}$ ($\text{GeV}^{-2}$)  \\ \hline
    $\rho $ & 13.49(45)  & 2.30(7) \\
    $\omega $ & 0  &  7.28(10) \\
    $\phi $ & 0   &  9.38(56) \\
    \hline
    \end{tabular}
\end{center}
\end{table}
One can now evaluate $k_V$ in Eq.~\eqref{eq:kNdef}. The results are given in Table~\ref{tab:3} for 11 values of $|t|$ below 1 GeV$^2$ at $\elab = 9$ GeV. 

We now turn our attention to the unnatural exchanges. 
The exchanges of $b$, $h$ and $h'$ contribute only to $A_2$ with 
\begin{align} 
A^{(\prime)A}_2(s,t) & = g^{(\prime)}_{A\gamma} g_{2 A} \frac{1- e^{-i\pi \alpha_A(t)}}{ \sin\pi \alpha_A(t)} s^{\alpha_A(t)-1}.
\end{align}
We consider only the $b$ and $h$ Regge poles in $A_2$. By neglecting the hidden strangeness exchange $h'$, the deviation of $\Sigma'/\Sigma$ from unity will be due to the $\phi$ exchange. 
It is empirically difficult to distinguish between $b$ and $h$ exchange. We will assume that they couple identically to the nucleon  $g_{2b} = g_{2h}$ and have degenerate trajectories $\alpha_b(t)  = \alpha_h(t)$. 

The Regge poles on the  $J^{PC} = (2,4,\ldots)^{--}$ trajectory contribute to $A_3$ only. This amplitude, which is determines the difference between target and recoil asymmetries at high energies, is known to be small for the similar reaction $\gamma p \to \pi^0 p$~\cite{Mathieu:2015eia}. Furthermore, the recent beam asymmetry measurements~\cite{AlGhoul:2017nbp} showed that $\Sigma \approx 1$, setting an upper limit to the $A_3$ contribution to $\eta$ photoproduction~\cite{Nys:2016vjz}.
Without any indication of significant $A_3$ contribution in $\gamma p \to \eta^{(\prime)} p$, we ignore it. These assumptions can be relaxed in the more flexible parametrization available online~\cite{JPACweb}. 
Under the above assumptions, one obtains
\begin{align}\label{eq:kUdef}
k_A & = \frac{\left| g'_{b\gamma} + g'_{h \gamma} \right|^2}{\left| g_{b\gamma} + g_{h \gamma} \right|^2}.
\end{align}

Ideally, one would determine the couplings $g_{A\gamma}^{(\prime)}$ following the procedure used in determination of the vector couplings.
Unfortunately, there is no data on axial-vector radiative decays $b,h\to \eta^{(\prime)} \gamma$. We can, however, estimate $k_A$ assuming that the axial-vector exchanges $b$ and $h$ follow the same pattern as the $\rho$ exchange:
\begin{align} \nonumber
k_A = k_\rho & = \frac{\Gamma (\eta' \to \gamma \rho)}{3\Gamma (\rho \to \gamma \eta)}  \left(\frac{m_{\eta'}}{ m_\rho} \cdot \frac{m_{\rho}^2-m_\eta^2}{ m_{\eta'}^2- m^2_\rho}\right)^3 \\
& = 0.685(59).
\label{eq:kU}
\end{align}
Alternatively one could use the value obtained from $\omega$ decay
\begin{align} \nonumber
k_\omega & = \frac{\Gamma (\eta' \to \gamma \omega)}{3\Gamma (\omega \to \gamma \eta)}  \left(\frac{m_{\eta'}}{ m_\omega} \cdot \frac{m_{\omega}^2-m_\eta^2}{ m_{\eta'}^2- m^2_\omega}\right)^3 \\
& = 0.884(99),
\label{eq:kUbis}
\end{align}
or a combination of the two. Based on vector-meson dominance (VMD) and SU(3) flavor symmetry, one can estimate the $b$ and $h$ relative couplings to $\eta^{(\prime)} \gamma$. Overall the isoscalar contribution is found to be suppressed by a factor of three  relative to the isovector contributions due to the $\gamma \eta^{(\prime)}$ vertex.
We therefore assume that $b$ is the dominant unnatural exchange. 

Using VMD and SU(3) flavor symmetry, it is therefore natural to assume that $k_A = k_\rho$. Without more information about axial-vector mesons, we consider $k_A$ constant in the forward direction. As an estimation of the systematic error, we investigated the effect on $\epsilon$ when $k_A  = k_\omega$ is used. 

\begin{figure}[htb]
\centerline{
\includegraphics[width=\linewidth]{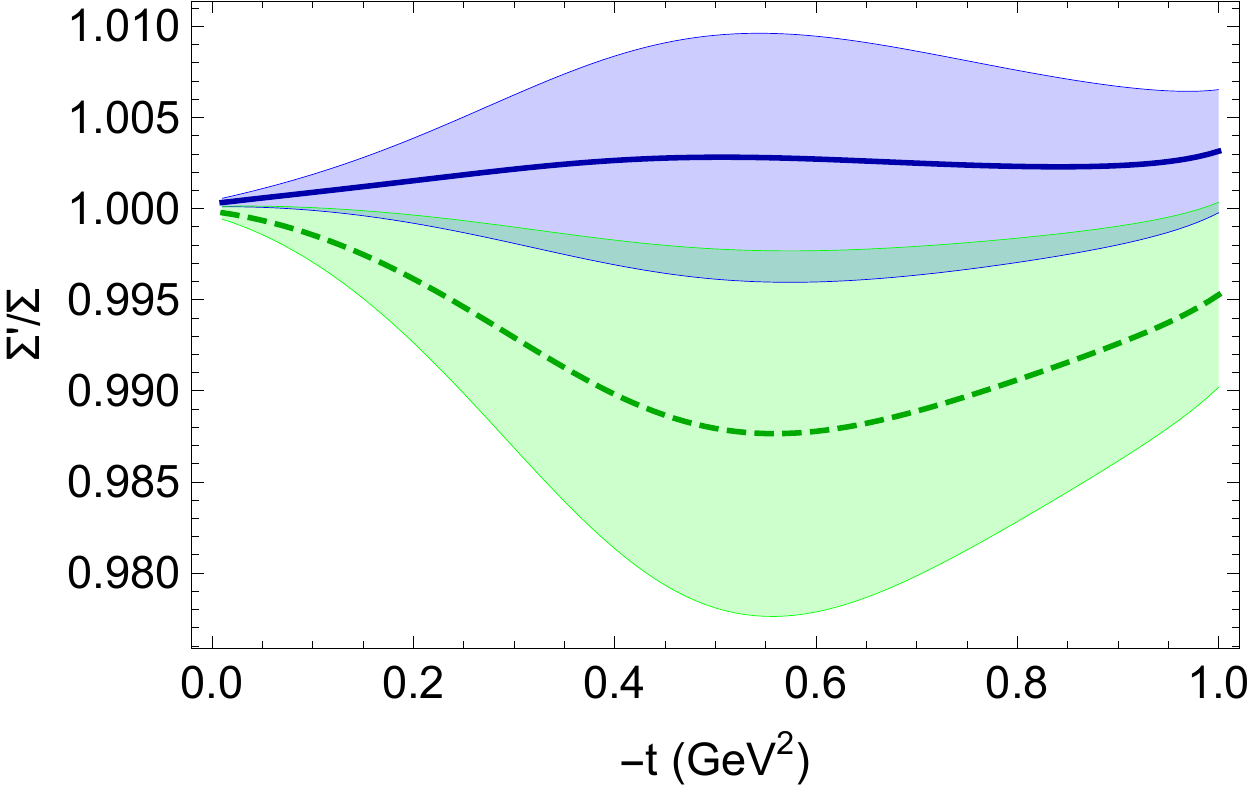}}
\caption{\label{fig:prediction} Predictions for the ratio of $\eta'$ and $\eta$ photoproduction beam asymmetries (dark blue line). The blue band illustrates the $1\sigma$ uncertainty on the prediction. The green dashed line depicts the predictions for $k_A=k_\omega$ in Eq.~\eqref{eq:kUbis}, with its corresponding $1\sigma$ uncertainty.}
\end{figure}

Because of this, the hidden strangeness exchange is given by the $\phi$ only. One has
\begin{align}\nonumber
k_\phi & = \frac{\Gamma (\phi \to \gamma \eta')} {\Gamma (\phi \to \gamma \eta)} 
\left( \frac{m_\phi^2 - m_\eta^2}{m_\phi^2 - m_{\eta'}^2} \right)^3 \\   & = 1.063(41).
\end{align}
Since $k_V > k_A = k_\rho$, one expects $\Sigma' > \Sigma$ and hence $\epsilon > 0$. Similarly, one expects $\epsilon < 0$ for $k_A=k_\omega$.
The only remaining unknown quantity needed to estimate the ratio $ \Sigma'/\Sigma$ in Eq.~\eqref{eq:ratio} is the $\eta$ 
beam asymmetry. One could use the recent GlueX
data~\cite{AlGhoul:2017nbp} as input. 
However, the analysis contains measurements at only 
four  values of $t$. We opt instead to use the predictions from~\cite{Nys:2016vjz},
which allows us to evaluate $\Sigma$ and $\epsilon$ 
in a variable $t$ range. 
This approach is justified by the observation 
that the prediction is in agreement 
with the GlueX measurements in~\cite{AlGhoul:2017nbp} 
and consistent with our hypotheses (negligible $h'$ pole, 
couplings proportional to decay widths and factorization 
of Regge poles). 
For completeness, we list the $\eta$ beam asymmetry from~\cite{Nys:2016vjz} in Table~\ref{tab:3} for 11 values of $t$ in the range $0 \leq -t/\text{GeV}^2 \leq 1 $. Our final result for $\Sigma'/\Sigma$ at $\elab = 9$ GeV is presented in Fig.~\ref{fig:prediction}. Note that we applied first-order error propagation to estimate the statistical error on the relevant quantities. We note, however, that the systematic errors coming from $k_A$ are larger than the statistical errors. This is illustrated in Fig.~\ref{fig:ratio}.

\begin{table}[htb]\caption{List of results for $k_V$ and $\epsilon$, where the latter is provided for both assumption $k_A = k_\rho$ and $k_A = k_\omega$. We also provide the input for the $\eta$ beam asymmetry from~\cite{Nys:2016vjz}. $t$ is expressed in GeV$^2$. \label{tab:3}}\begin{center}
    \begin{tabular}{|c|c|c|c|c|}
    \hline
    $t$& $\Sigma$ & $k_V$ & $\epsilon \times 10^4$ &  $\epsilon \times 10^4$\\
     &   &   & $k_A = k_\rho$ & $k_A = k_\omega$\\\hline
    -0.1 & 0.990  & 0.756(49)  & 10(10) & -18(16)\\
    -0.2 & 0.977  & 0.737(51) & 17(24) & -47(37)\\
    -0.3 & 0.961  & 0.728(52) & 23(42) & -85(64)\\
   -0.4 & 0.946  & 0.722(53) &  28(59) & -123(89)\\
   -0.5 &  0.938 & 0.719(53) &  30(69) & -145(104)\\
   -0.6 & 0.938  & 0.717(54) &  29(69) & -147(104)\\
   -0.7 & 0.944  & 0.718(54) &  26(63) & -134(95)\\
   -0.8 & 0.951  & 0.720(54) &  24(54) & -114(82)\\
   -0.9 & 0.959  & 0.728(53) &  25(45) & -90(68)\\
   -1.0 & 0.965  & 0.756(50) &  33(35) & -60(53)\\    
	\hline
    \end{tabular}
\end{center}
\end{table}

From Table~\ref{tab:3} and Fig.~\ref{fig:prediction}, one observes that the quantity $\epsilon = \Sigma'/\Sigma-1 $ is predicted to be small, of the order $10^{-3} - 10^{-4}$ for $k_A = k_\rho$.  This is expected due to the presence of the factor $1-\Sigma^2$ in Eq.~\eqref{eq:ratio}. Changing the value of $g_{4\phi}$ (e.g.\ by factor of $2$) does not have a notable effect on this conclusion.
The prediction with $k_A = k_\omega$ in green on Fig.~\ref{fig:prediction} is indicated only to illustrate  sensitivity to model assumptions. As we argued above, the value $k_A = k_\rho$ is favored by $SU(3)$ and the quark model.  
%A sizable deviation of $\epsilon$ from unity would indicate one or more of the following possibilities: $i)$ the axial-vector meson radiative decay is significantly different from the quark model predictions;  $ii)$ the hidden strangeness $h'$ Regge pole is not negligible;$iii)$ the $\phi$ exchange has been incorrectly estimated; $iv)$ the contribution of exchanges in the $A_3^{(\prime)}$ amplitudes are significant. 

Even though the level of precision to resolve a difference between $\eta$ and $\eta'$ photoproduction beam asymmetries might not be achieved with the GlueX or CLAS12 detectors,  
%  Without sufficient resolution, we predict the same beam asymmetry for $\eta$ as for %$\eta'$ production. 
we remark that $\Sigma$ decreases as $|t|$ increases, resulting in  $\epsilon$ increasing with $|t|$. This trend might nevertheless be observable at the JLab facility. When the measurements will be available, the reader can test all these different hypotheses independently by playing with a flexible parametrization on the JPAC website~\cite{JPACweb, Mathieu:2016mcy}. 
However, given that the estimate is based on rather reasonable assumptions, a significant deviation from our prediction, if observed at these experiments, would  require a new approach of meson photoproduction.

%indicate potential for ``new physics'' needed to describe meson photoproduction.

In our calculation, we proceeded by a separate evaluation of the natural ($k_V$) and unnatural ($k_A$) exchanges to photoproduction. Note that the quantity $\Sigma'/\Sigma$ is only mildly sensitive to the precise value of $k_A$, since the dynamics are dominated by natural exchanges. This is illustrated in Fig.~\ref{fig:prediction} where we plot the result for $k_A=k_\omega$.
Separate measurements of $k_V$ and $k_A$, given by Eq.~\eqref{eq:kNkU}, would provide useful information to identify the source of deviations. In particular, it would provide us with more detailed information on the coupling of $b$ and $h$ to $\eta'\gamma$ relative to their coupling to $\eta\gamma$ (see the discussion related to Eq.~\eqref{eq:kU}). These couplings are experimentally unconstrained at the moment.

\section*{Acknowledgments}
We thank J.~Stevens for referring us to the GlueX measurement and N. Sherrill for comments on the manuscript.
This material is based upon work supported in part by the U.S.~Department of Energy, Office of Science, 
Office of Nuclear Physics under contract DE-AC05-06OR23177. 
This work was also supported in part by the U.S.~Department of Energy under Grant DE-FG0287ER40365, 
National Science Foundation under Grants PHY-1415459 and PHY-1205019, the IU Collaborative Research Grant and the Research Foundation Flanders (FWO-Flanders).
CF-R work is supported in part by PAPIIT-DGAPA (UNAM) grant No.~IA101717,
by CONACYT (Mexico) grant No.~251817, 
and by Red Tem\'atica CONACYT de F\'{\i}sica en Altas Energ\'{\i}as (Red FAE, Mexico). 

\bibliographystyle{elsarticle-modified}
\bibliography{quattro}
\end{document}